\documentclass[preprint,showpacs,preprintnumbers,amsmath,amssymb,floats,nofootinbib]{revtex4}
\usepackage[dvips]{graphicx}
\usepackage[english]{babel}
\usepackage{amsmath}
\usepackage{amssymb}
\usepackage{bm}

\newcommand{\vk}{\mathbf{k}}

\newcommand{\vR}{\mathbf{R}}

\newcommand{\be}{\begin{eqnarray}}
\newcommand{\ee}{\end{eqnarray}}
\newcommand{\p}{\partial}

\def\ket#1{|#1\rangle}
\def\bra#1{\langle #1 |}
\def\ep#1{\langle #1 \rangle}

\begin{document}

\title{Berry phase and Anomalous Hall Effect in a Three-orbital Tight-binding Hamiltonian}
\author{Yan He$^1$, Joel Moore$^{2,3}$ and C.M. Varma$^1$}

\affiliation{$^1$ Department of Physics, University of California, Riverside, CA\\
$^2$Department of Physics, University of California, Berkeley, CA \\
$^3$Materials Sciences Division, Lawrence Berkeley Laboratory, Berkeley, CA}
\date{\today}
\begin{abstract}
We consider the Anomalous Hall (AH) state induced by interactions in a three-orbital per unit-cell model. To be specific we consider a model appropriate for the Copper-Oxide lattice to highlight the necessary conditions for time-reversal breaking states which are AH states and which are not.  We compare the singularities of the wave-functions of the three-orbital model, which are related to the nonzero Berry curvature, and their variation with a change of gauge to those in the two-orbital model introduced in a seminal paper by Haldane. Explicit derivation using wave-functions rather than the more powerful abstract methods may provide additional physical understanding of the phenomena.
\end{abstract}

\maketitle

\section{Introduction}

\begin{figure}
\centerline{\includegraphics[width=0.5\textwidth]{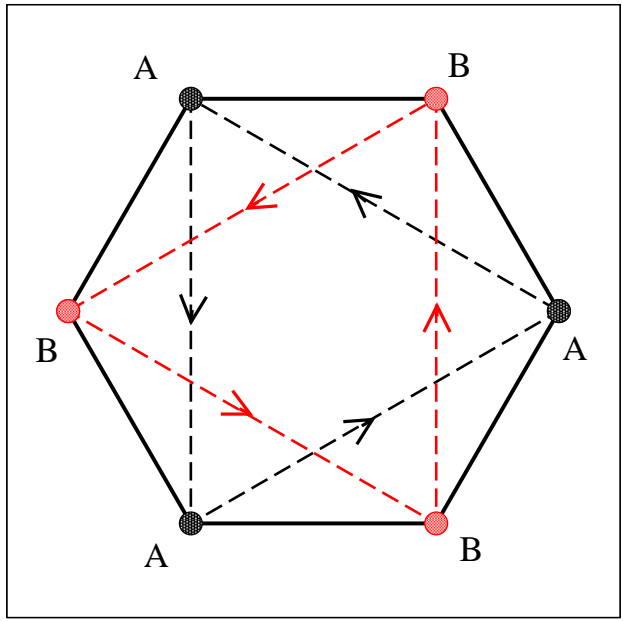}}
\caption{Schematic figure of loop current in hexagon lattice}
\label{hexagon}
\end{figure}

In asking the question whether a quantized Hall effect may exist, in principle, without an applied magnetic field, Haldane \cite{haldane} introduced an effective one-electron model on the two-orbital per unit-cell hexagonal crystal with complex next nearest transfer integrals. This has turned out to be a fecund contribution. It further augmented the topological arguments of Thouless and collaborators \cite{tknn}  for the quantum Hall effect. It also introduced the general discussion of the topological features of band-structures in two-dimensions and in particular led through further imaginative work to the suggestion and discovery of topological insulators \cite{kane-mele,hasankane,moorenature,qizhangreview}. Haldane's model also showed the connection of topological properties to the time-reversal-violating states due to orbital current loops in the lattice without changing translational invariance, see Fig. (\ref{hexagon}).

Orbital current loops without changing translational symmetry were predicted to arise as broken symmetry states due to interactions in a three-orbital per unit cell model for underdoped cuprates \cite{cmv97, simon-cmv, cmv06} and have been discovered in several families of cuprates \cite{bourges, kaminski, li}. These loop-current states, however, do not lead to the quantized anomalous Hall effect (QAHE) or ``Chern insulator'' discussed by Haldane. The difference is that the loop-current states violate both time-reversal ${\cal R}$ and inversion ${\cal I}$ but preserve their product ${\cal I}$. It was pointed out by Fradkin and Sun \cite{sun-fradkin} that there is no QAHE effect when ${\cal RI}$ symmetry is present; they also pointed out that a QAHE state for the three-orbital cuprate model is possible, in principle through a state with symmetry different from that observed.

In this paper, we elaborate on the work of Fradkin and Sun by deriving the Berry phase responsible for the QAHE state and also how the singularities of the wavefunctions vary with the choice of gauge. When the topologically active band (i.e., the one with nonzero Chern number or TKNN integer~\cite{tknn}) is partially rather than completely filled, the Berry phase will still contribute to a non-quantized intrinsic anomalous Hall effect~\cite{karplusluttinger,niuahe,haldaneahe,nagaosaahereview}.  For comparison, we also rederive wavefunctions for the two-orbital case.  The physics of the three-orbital model is not surprising in the context of previous general work on topological effects in non-interacting lattice models; we hope that we have discussed it in a new fashion which will be useful to non-experts and that the particular model explains how topological effects might appear in three-band materials.  Whenever possible we seek to explain the value of a Berry-phase calculation geometrically rather than simply stating the result.  We also work with explicit wavefunctions rather than more powerful abstract methods, although a few comments are provided regarding the latter.

\section{Anomalous Hall states in the Cu-O model}

\begin{figure}
\centerline{\includegraphics[width=0.5\textwidth]{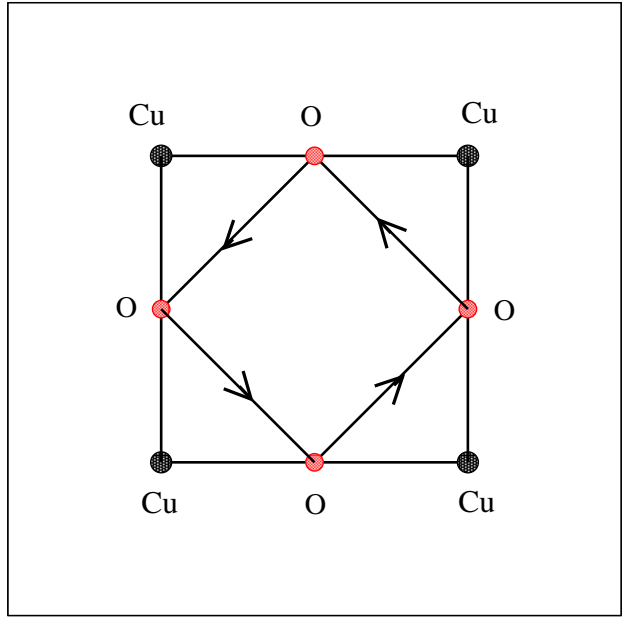}}
\caption{Schematic figure of loop current in Cu-O lattice}
\label{fig:cuo}
\end{figure}

Consider the two-dimensional lattice with the structure of the copper oxides, Fig. (\ref{fig:cuo}). There are three orbitals per unit-cell, the $d$-orbital on the copper atom and the $p_x$ and $p_y$ orbitals on the oxygens.  The minimal kinetic energy operator with a choice of gauge such that the $d$ orbital is purely real and the $p_x$ and the $p_y$ orbitals purely imaginary is
\be
\label{ke}
H_{KE}=it\,d^{\dagger}_{\vk}(s_xp_{x,\vk}+s_yp_{y,\vk})
-t's_xs_yp_{x,\vk}^{\dagger}p_{y,\vk}+h.c.
\ee
with $s_x=\sin(k_x/2)$ and $s_y=\sin(k_y/2)$ for a lattice constant taken to be 1. For simplicity, let the fermions be spinless.
Consider only the interaction between the $p$ orbitals,
\be
\label{int}
H_{int}=\sum_{\ep{i,j}}Vn_{p,i}n_{p,j}.
\ee
Following the procedure with which some time-reversal violating states were derived for the cuprates, we use the operator identity (for spin-less fermions),
\be
\label{op-id}
n_i n_j = -\frac{1}{2} \big( |j_{ij}|^2 + n_i + n_j\big),
\ee
where $j_{ij}$ is the self-adjoint operator
\be
j_{ij} = i(c_i^{\dagger}c_j - c_j^{\dagger}c_i).
\ee

Decomposing\footnote{In general the diagonal in spin indices part of $n_in_j$ gives (\ref{op-id}), the non-diagonal can be written in terms of products of spin-currents with which mean-field theory gives the possibility of symmetry breaking topological spin-current states. The decompositions of the operator may also be done without the ``i" in (\ref{op-id}) from which various Pomeranchuk instabilities may be derived for symmetry breaking in various  irreducible representations for spin and charge densities.} the interaction term in (\ref{int}), by defining
\be
(V/2) \langle j_{ij} \rangle = i r,
\ee
 and doing a mean field calculation, one finds an additional kinetic energy term
\be
H_{int}'=irc_xc_yp_{x,\vk}^{\dagger}p_{y,\vk}+h.c.
\ee
If $r\neq0$ is a stable state, it describes loop currents flowing clockwise (or anti-clockwise) around the oxygen's in each unit-cell as shown in Fig (\ref{fig:cuo}). This is one of the five possible loop-curent states with non-overlapping loops in the Cu-O lattice all of which preserve translational symmetry \cite{aji-shekhter-cmv}). In (\ref{fig:cuo}), the flux has one sign in the square formed by the
nearest neighbor oxygens which surround a cu and another sign in the square formed by the
nearest neighbor oxygens which do not surround a cu. As pointed out by Fradkin and Sun \cite{sun-fradkin} such a time-reversal violating state, which does not change translational symmetry or break inversion symmetry, satisfies all the conditions of a Haldane state for the Cu-O lattice. The other four loop-current states do not. One of those is just the photon on a lattice and cannot order. The other three can order and indeed order consistent with the symmetry of two of them (in different domains) is observed in under-doped cuprates. So, our consideration of states such as in (\ref{fig:cuo}) is only a specific example to illustrate the nature of AH states in 3-orbital models.

We will consider the Haldane state (quantized anomalous Hall effect) of the cu-o model and therefore the singularities of the model with the Hamiltonian $H = H_{KE} + H_{int}'$. Before we do that, let us consider the simpler case of two-orbitals per unit-cell.

\section{Two-band models}

A general Hamiltonian in the space of two orbitals per unit-cell may be written as, ignoring an overall shift of the energy that does not affect the Berry phase and assuming that there is no basis of the Bravais lattice,
\be
H={\bf R}({\bf k})\cdot\bm{\sigma}=\left(\begin{array}{cc}
R_3 & R_1-iR_2\\
R_1+iR_2 & -R_3
\end{array}
\right)
\ee
Here $R_i$ for $i=1,2,3$ are some smooth functions of $k_x$ and $k_y$ with period $2\pi$. For now, we do not need the detailed form of these functions. It is easy to diagonalize the above Hamiltonian to find that there are two bands, $E=\pm R$, with $R=\sqrt{R_1^2+R_2^2+R_3^2}$.

Consider the lower band, $E=-R({\bf k})$. The eigenstate can be written in two ways
corresponding to two different choices of gauge; (the point of studying apparent consequences of the choice of gauge will be clear in a later section of the paper):
\be
\ket{\psi^{A}}=\frac{1}{\sqrt{2R(R-R_3)}}\left(\begin{array}{c}
R_3-R\\
R_1+iR_2
\end{array}\right)\\
\ket{\psi^{B}}=\frac{1}{\sqrt{2R(R+R_3)}}\left(\begin{array}{c}
R_1-iR_2\\
-R-R_3
\end{array}\right)
\ee
They are connected by a $U(1)$ gauge transformation
\be
\ket{\psi^{B}}=e^{i\phi}\ket{\psi^{A}},\quad\mbox{with}\,\,
e^{\phi}=\frac{-R_1+iR_2}{\sqrt{R_1^2+R_2^2}},\quad\phi=-\arctan(R_2/R_1)
\ee
Then the {\it Berry phase}, $A_{\mu}$  is also gauge dependent, given for the two choices respectively by
\be
&&A^{A}_{\mu}\equiv-i\bra{\psi^{A}}\nabla_{\mu}\ket{\psi^{A}}
=-\frac{1}{2R(R-R_3)}\Big(R_2\frac{\p R_1}{\p k_{\mu}}-R_1\frac{\p R_2}{\p k_{\mu}}\Big)\\
&&A^{B}_{\mu}\equiv-i\bra{\psi^{B}}\nabla_{\mu}\ket{\psi^{B}}
=\frac{1}{2R(R+R_3)}\Big(R_2\frac{\p R_1}{\p k_{\mu}}-R_1\frac{\p R_2}{\p k_{\mu}}\Big).
\ee
$A_{\mu}$'s are also connected by a $U(1)$ gauge transformation:
\be
A^{B}_{\mu}=A^{A}_{\mu}+\nabla_{\mu}\phi,\quad
\nabla_{\mu}\phi=-\frac{1}{R^2-R_3^2}
\Big(R_2\frac{\p R_1}{\p k_{\mu}}-R_1\frac{\p R_2}{\p k_{\mu}}\Big)
\ee
The {\it Berry curvature} is gauge invariant and given by
\be
F_{xy}=\frac{\p A_y}{\p k_x}-\frac{\p A_x}{\p k_y}
&=&\frac1{2R^3}\epsilon_{abc}R_a\frac{\p R_b}{\p k_x}\frac{\p R_c}{\p k_y}
=\frac12\epsilon_{abc}\hat{R}_a\frac{\p \hat{R}_b}{\p k_x}
\frac{\p \hat{R}_c}{\p k_y}
\ee
Here $\hat{\bf R}={\bf R}/R$ is a unit vector. If we integrate over the entire Brillouin zone, we find
\be
c=\frac{1}{2\pi}\int d^2k F_{xy}=\frac{1}{4\pi}\int d^2k
\epsilon_{abc}\hat{R}_a\frac{\p \hat{R}_b}{\p k_x}\frac{\p \hat{R}_c}{\p k_y}
\ee

This is the well known result \cite{qi-wu} that the Chern number of a two band model is equivalent to the winding number of the mapping from a 2D Brillouin zone which is 2D torus($T^2$) to the 2D unit sphere($S^2$). This mapping can be understood by taking spherical coordinates $\vR=R(\sin\theta\cos\phi,\sin\theta\sin\phi,\cos\theta)$. The wave function can be written as $\ket{\phi}=(-\sin(\theta/2),\cos(\theta/2)e^{i\phi})$ which is a two component spinor. The unit sphere is just the Bloch sphere $\hat{\vR}=-\bra{\phi}\bm{\sigma}\ket{\phi}$ associated with this spinor.

The Berry phase ${\bf A}$ is a vector field defined on the momentum space or the torus. Since the torus is mapped to a sphere, one can also think of the Berry phase as defined on this sphere. In spherical coordinates, we have
\be
{\bf A}^A=\frac{1+\cos\theta}{2\sin\theta}\hat{\bm{\phi}},\quad
{\bf A}^B=-\frac{1-\cos\theta}{2\sin\theta}\hat{\bm{\phi}}
\ee
This is just the vector potential of the Wu-Yang monopole \cite{nakahara} on a unit sphere. The magnetic field is ${\bf B}=\nabla\times{\bf A}
=\frac{1}{\sin\theta}\frac{\p}{\p\theta}(A_{\phi}\sin\theta)\hat{\bf r}=-\frac12\hat{\bf r}$. The magnetic field is like that of a monopole with charge $g=-2\pi$. Furthermore, this implies that a monopole-like singular point is located at the center of the sphere $R_1=R_2=R_3=0$, which is also the point at which the two bands become degenerate.

The winding number is easier to compute than the Chern number, since it is directly written in terms of matrix elements of the Hamiltonian and does not require computing the eigenvectors. If $R_1$, $R_2$ and $R_3$ are independent of each other, and each of them can take both positive and negative values at the point where other two components are zero, then $\hat{{\bf R}}$ will sweep out the whole unit sphere which encloses the singular point inside it. Then the winding number or the Chern number is nontrivial. But in general it is also possible that $\hat{{\bf R}}$ winds the sphere twice in opposite directions and cancel out or comes back to a given point as ${\bf k}$ is varied over all its values without  sweeping the entire sphere. Then the Chern number is zero.

As a specific example \cite{zhang}, take $\vR=(\sin k_x,\sin k_y,m+\cos k_x+\cos k_y)$. For $0<m<2$, there is always a gap between the two bands and it is easy to see that $\hat{\vR}$ sweeps the whole sphere. One can also directly verify that
\be
c=\frac{1}{4\pi}\int d^2k\frac{\cos k_x+\cos k_y+m\cos k_x\cos k_y}
{[\sin^2 k_x+\sin^2 k_y+(m+\cos k_x+\cos k_y)^2]^{3/2}}=-1
\ee
Consider next the Haldane model \cite{haldane}. It has a staggered flux inside each unit cell. The Hamiltonian is
\be
H=t_1\Big(\sum_i\cos(\vk\cdot{\bf a}_i)\sigma^1
+\sum_i\sin(\vk\cdot{\bf a}_i)\sigma^2\Big)
+\Big(M-2t_2\sin\phi\sum_i\sin(\vk\cdot{\bf b}_i)\Big)\sigma^3
\ee
The maximum of $\sum_i\sin(\vk\cdot{\bf b}_i)$ is $3\sqrt{3}/2$, thus for $|M/t_2|<3\sqrt{3}|\sin\phi|$, $\hat{\vR}$ will cover the whole sphere. Indeed, for $\phi>0$, one can directly verify that $c=1$ in this case.

The condition on $M$ is simply the necessary condition for the AH state that the monopole singularity exist which requires that $R_3({\bf k})$ go through zero at some point ${\bf k}$ and change sign as that point is crossed.

Fradkin and Sun \cite{sun-fradkin}  pointed out that if both time-reversal $\mathcal {R}$ and inversion $\mathcal{I}$ are broken but the product $\mathcal{R} \mathcal{I}$ conserved, there can be no anomalous Hall state: the diagonal components of the Berry vector potential must vanish.  Note that $\mathcal{R} \mathcal{I}~ R_3\sigma_3 = -R_3 \sigma_3$. Therefore if $\mathcal{R} \mathcal{I}$ is conserved $R_3({\bf k}) = 0$ for all ${\bf k}$.  Then the sphere $\hat{R}({\bf k})$ turns to a circle and no singularity can be defined.

In contrast to this is the $\theta_{II}$ type loop current states realized in the three-orbital model for cuprates \cite{bourges, kaminski, li}. Its mean field Hamiltonian in the same basis as (\ref{ke}) is
\be
H=\left(\begin{array}{ccc}
0 & its_x+irc_x & its_y+irc_y\\
-its_x-irc_x & 0 & t's_xs_y\\
-its_y-irc_y & t's_xs_y & 0
\end{array}\right)
\ee
 If we define $id^{\dagger}=\tilde{d}^{\dagger}$, then the Hamiltonian in the new basis is a real matrix. Actually, if $\mathcal{R} \mathcal{I}$ is invariant, then the Hamiltonian (in momentum space) can always be written as a real matrix and the phase of all eigenvectors are constant. Therefore such a state though violating time-reversal cannot have an AH state.

\section{Three band Loop current model with AH}

\subsection{Chern number and winding number}
Now we turn back to the three-orbital copper-oxygen model given by Hamiltonian $H=H_{KE} + H' $, which in the space of $d, p_x, p_y$ is
\be
H=\left(\begin{array}{ccc}
0 & its_x & its_y\\
-its_x & 0 & t's_xs_y+irc_xc_y\\
-its_y & t's_xs_y-irc_xc_y & 0
\end{array}\right)
\ee
with $s_x=\sin(k_x/2)$ and $c_x=\cos(k_x/2)$, etc. Note that even though the matrix elements as a function of $k_x$ and $k_y$ do not have the  period $2\pi$, the energy dispersion as a function of $k_x$ and $k_y$ does have a period $2\pi$.  As commented on below, the Hamiltonian for a tight-binding model with a basis is not strictly periodic when ${\bf k}$ is translated by a reciprocal lattice vector ${\bf G}$, but rather is transformed by the unitary matrix $\exp(i {\bf G} \cdot {\bf a}_i$, where ${\bf a}_i$ is the location of the $i$th site in the unit cell.

To diagonalize the above Hamiltonian, one has to solve a cubic equation to find out the eigenvalues. Since Chern number is topological invariant, if we deform the Hamiltonian without the bands crossing, the Chern number will stay the same. We can therefore simplify the problem by dropping the $t's_xs_y$ term and come back later to ensure that this simplification is valid. The simplified Hamiltonian is
\be
H'=\left(\begin{array}{ccc}
0 & its_x & its_y\\
-its_x & 0 & irc_xc_y\\
-its_y & -irc_xc_y & 0
\end{array}\right)
\ee
This Hamiltonian can written as $H={\bf R}\cdot{\bf L}$ with $R_1=ts_x$, $R_2=-ts_y$, $R_3=rc_xc_y$ and $L_x$, $L_y$, $L_z$ are the spin 1 representation of the $SU(2)$ generators in contrast to the spin 1/2 representation of the $SU(2)$ generators for the two band case.

We will discuss the location of singular points in simple gauges for model $H'$ of Eq (4). They should be qualitatively similar to model $H$ of Eq (3). It is easy to find the eigenvalues,  $E=0,\pm R$ with $R=\sqrt{R_1^2+R_2^2+R_3^2}$.
 Let us focus on the lowest band $E_1=-R$, the corresponding eigenvector can be written in two different ways labeled by $A,B$,
\be
&&\ket{\psi_1^A}=\frac{1}{\sqrt{2(R^2-R_3^2)}R}\Big(R^2-R_3^2,\,\,
-R_2R_3+iRR_1,\,\,-R_1R_3-iRR_2\Big)^T\\
&&\ket{\psi_1^B}=\frac{1}{\sqrt{2(R^2-R_1^2)}R}\Big(-R_1R_3+iRR_2,
\,\,-R_1R_2-iRR_3,\,\,R^2-R_1^2\Big)^T
\ee
These two wave functions are related by a $U(1)$ gauge transformation
\be
\ket{\psi_1^B}=e^{i\phi}\ket{\psi^{A}},\quad e^{i\phi}=\frac{-R_1R_3+iRR_2}{\sqrt{(R^2-R_1^2)(R^2-R_3^2)}},\quad
\phi=-\arctan\Big(\frac{RR_2}{R_1R_3}\Big)
\label{u1}
\ee
The Berry phase is given by
\be
&&A^{A}_{\mu}=-i\bra{\psi^{A}}\nabla_{\mu}\ket{\psi^{A}}
=-\frac{R_3}{R(R^2-R^2_3)}\Big(R_2\frac{\p R_1}{\p k_{\mu}}-R_1\frac{\p R_2}{\p k_{\mu}}\Big)\\
&&A^{B}_{\mu}=-i\bra{\psi^{B}}\nabla_{\mu}\ket{\psi^{B}}
=-\frac{R_1}{R(R^2-R^2_1)}\Big(R_3\frac{\p R_2}{\p k_{\mu}}-R_2\frac{\p R_3}{\p k_{\mu}}\Big)
\label{vector}
\ee
They are also connected by a $U(1)$ gauge transformation,
\be
A^{B}_{\mu}=A^{A}_{\mu}+\nabla_{\mu}\phi
\ee
The Berry curvature is given by
\be
F_{xy}=\frac{\p A_y}{\p k_x}-\frac{\p A_x}{\p k_y}
&=&-\frac1{R^3}\epsilon_{abc}R_a\frac{\p R_b}{\p k_x}\frac{\p R_c}{\p k_y}
=-\epsilon_{abc}\hat{R}_a\frac{\p \hat{R}_b}{\p k_x}
\frac{\p \hat{R}_c}{\p k_y}
\ee
Note there is a factor of 2 difference compared to the two-band model. The Chern number is the integral of the Berry curvature in the BZ
\be
c&=&\frac1{2\pi}\int d^2kF_{xy}
=-\frac1{2\pi}\int d^2k\epsilon_{abc}\hat{R}_a\frac{\p \hat{R}_b}{\p k_x}
\frac{\p \hat{R}_c}{\p k_y}\nonumber\\
&=&-\frac1{2\pi}\int d^2k\frac{t^2r(c_x^2+s_x^2c_y^2)}{4[(ts_x)^2+(ts_y)^2+(rc_xc_y)^2]^{3/2}}
\ee
In this model, $R_3=rc_xc_y$  takes only positive values and $R_1=ts_x$ and $R_2=-ts_y$ continuously vary from $-t$ to $t$ and do not go back. Thus $\hat{\vR}$ only sweeps half of the unit sphere. Thus the surface integral
$\int d^2k\hat{\vR}(\p_x\hat{\vR}\times\p_y\hat{\vR})=2\pi$ and Chern number $c=-1$.

We have obtained the relation between the Chern number and winding number for the spin 1/2 and 1 representations of $SU(2)$ algebra. Actually, we can generalize this relation to any spin $n/2$ representation of $SU(2)$ for integer $n$. The Hamiltonian is given by
\be
H=R_1J_x+R_2J_y+R_3J_z
\ee
Here $J_a$ for $a=x,y,z$ are $n$ by $n$ matrices and satisfy $[J_a,J_b]=i\epsilon_{abc}J_c$. Then the eigenvalues are $E_i/R=-n/2,-(n/2-1),\cdots,n/2-1,n/2$. Then for the $i$th band, we have
\be
c=\frac1{2\pi}\int d^2kF_{xy}
=\frac{E_i}{R}\cdot\frac1{2\pi}\int d^2k\epsilon_{abc}\hat{R}_a\frac{\p \hat{R}_b}{\p k_x}
\frac{\p \hat{R}_c}{\p k_y}
\ee
To prove the above result, it is better to consider $H/R=\hat{\vR}\cdot{\bf J}$ which has the same Chern number as $H$. Since $\hat{\vR}$ is a unit vector, it can be parameterized by   spherical coordinates $\hat{\vR}=(\sin\theta\cos\phi,\,\sin\theta\sin\phi,\,\cos\theta)$. Here $\theta$, $\phi$ are functions of $k_x$ and $k_y$.

To be specific, we take $J_z$ as a diagonal matrix $J_z=\mbox{diag}\{n/2,(n/2-1),\cdots,-n/2-1,-n/2\}$. We can make a rotation to diagonalize $H/R$ as,
\be
e^{i\theta J_y}e^{i\phi J_z}\hat{\vR}\cdot{\bf J}e^{-i\phi J_z}e^{-i\theta J_y}
=J_z
\ee
For the $i$th band, we have the eigenvector
\be
\psi_i=e^{-i\phi J_z}e^{-i\theta J_y}\mathbf{n}_i,\quad
\mathbf{n}_i(0,\cdots,1,\cdots,0)^T
\ee
In the above vector, only $i$th component is $1$ and all others are zero. The Chern number can
also be written in terms of differential forms as
\be
c=-\frac{i}{2\pi}\int d^2k\Big(\frac{\p\psi^{\dagger}}{\p k_x}\frac{\p\psi}{\p k_y}
-\frac{\p\psi^{\dagger}}{\p k_y}\frac{\p\psi}{\p k_x}\Big)
\equiv -\frac{i}{2\pi}\int d\psi^{\dagger}\wedge d\psi
\label{chern}
\ee
It is easy to find
\be
d\psi&=&\Big(-iJ_ze^{-i\phi J_z}e^{-i\theta J_y}d\phi
-ie^{-i\phi J_z}e^{-i\theta J_y}J_yd\theta\Big)\mathbf{n}_i\nonumber\\
d\psi^{\dagger}&=&\mathbf{n}_i^T\Big(ie^{i\theta J_y}e^{i\phi J_z}J_zd\phi
+iJ_ye^{i\theta J_y}e^{i\phi J_z}d\theta\Big)\nonumber
\ee
Using the above in Eq (\ref{chern}), we find
\be
c&=&-\frac{i}{2\pi}\int \mathbf{n}_i^T \Big(J_ye^{i\theta J_y}
J_ze^{-i\theta J_y}d\theta\wedge d\phi
+e^{i\theta J_y}J_ze^{-i\theta J_y}J_yd\phi\wedge d\theta\Big)\mathbf{n}_i\nonumber\\
&=&-\frac{i}{2\pi}\int\mbox{Tr}\Big[e^{i\theta J_y}(J_yJ_z-J_zJ_y)e^{-i\theta J_y}\cdot
(\mathbf{n}_i\mathbf{n}_i^T)\Big]d\theta\wedge d\phi\nonumber\\
&=&\frac{1}{2\pi}\int\mbox{Tr}\Big[e^{i\theta J_y}J_xe^{-i\theta J_y}\cdot
(\mathbf{n}_i\mathbf{n}_i^T)\Big]d\theta\wedge d\phi\nonumber\\
&=&\frac{1}{2\pi}\int\mbox{Tr}\Big[(\cos\theta J_x+\sin\theta J_z)\cdot
(\mathbf{n}_i\mathbf{n}_i^T)\Big]d\theta\wedge d\phi
\label{c2}
\ee
The matrix $(\mathbf{n}_i\mathbf{n}_i^T)$ has $1$ as its $i$th diagonal element, and all other elements are zero. Thus for any $n$ by $n$ matrix $A$, we have Tr$[A(\mathbf{n}_i\mathbf{n}_i^T)]=A_{ii}$. Since $J_x$ only has off-diagonal elements, only the second term of Eq(\ref{c2}) contributes. Therefore,
\be
c=\frac{1}{2\pi}(J_z)_{ii}\int\sin\theta d\theta\wedge d\phi
=\frac{E_i}{R}\frac{1}{2\pi}\int\sin\theta d\theta\wedge d\phi
\ee
On the other hand, the winding number integral in terms of spherical coordinates can be written as
\be
\int d^2k\epsilon_{abc}\hat{R}_a\frac{\p \hat{R}_b}{\p k_x}
\frac{\p \hat{R}_c}{\p k_y}=\frac12\int\epsilon_{abc}\hat{R}_a dR_b\wedge dR_c
=\int\sin\theta d\theta\wedge d\phi
\ee
Combining the above two equations, we find the desired results.

There is a subtle point about this loop current model's behavior when ${\bf k}$ is shifted by a lattice vector. Recall that $R_1=t\sin(k_x/2)$, $R_2=-t\sin(k_y/2)$, $R_3=r\cos(k_x/2)\cos(k_y/2)$ and they are functions of $k_x$ and $k_y$ with period $4\pi$ not $2\pi$. Thus the Hamiltonian is not invariant under changes $k_x\to k_x+2\pi$ and $k_y\to k_y+2\pi$. This happens generally if the lattice contains a basis. It is easy to see that the eigenstates are also not invariant under $2\pi$ shift. But if we rewrite the wave function in real space, we find
\be
\psi_1^A\propto\sum_ie^{i\vk\cdot{\bf r}_i}
\left(\begin{array}{c}
R^2-R_3^2\\
-(\frac{tr}{2}\sin k_y-itR)\sin(k_x/2)e^{ik_x/2}\\
-(\frac{tr}{2}\sin k_x+itR)\sin(k_y/2)e^{ik_y/2}
\end{array}\right)
\ee
Since $R$ and $R_{1,2,3}^2$ has period $2\pi$, the above wave function is invariant under $2\pi$ shift as one expected.

If we define the torus to be $-\pi<k_{x,y}<\pi$, then $R_3$ is always positive and the torus is mapped to half of a unit sphere by $\hat{\vR}$. It might seem that in this case, the monopole singular point is not enclosed by the half surface, therefore the Chern number need not to be quantized and  may have no topological meaning. But if we look at the vector potential in Eq.(\ref{vector}), they are invariant under $2\pi$ shift, and thus are well defined on the whole torus. Therefore, according to Dirac's arguments, the integral of field strength on this closed surface should give a quantized topological invariant. In the mapping $\hat{\vR}(k_x,k_y)$, the boundary of BZ is mapped to the equator of the sphere. As discussed above, the wave functions are the same at the two points like $(-\pi,k_y)$ and $(\pi,k_y)$ on the boundary of BZ. These two points are mapped to $(-\frac{1}{1+s_y^2},s_y,0)$ and $(\frac{1}{1+s_y^2},s_y,0)$ on the equator. These two points should be identified, and then the boundary of the half sphere is glued together to make a closed surface which is topologically equivalent to a sphere. In this loop current model, we have wound over this closed surface once.

\subsection{Singular points of wavefunctions and their gauge-dependence}

We now discuss the singular points of wave function inside the BZ for the three-orbital case. One can see that $\psi_1^A$ is well defined for all possible values of $R_{1,2,3}$ except when $R_3=\pm R$ or $R_1=R_2=0$. So this is the singular point of $\psi_1^A$. It corresponds to $k_x=k_y=0$ in the BZ. Similarly, $\psi_1^B$ has singular points when $R_1=\pm R$ or $R_2=R_3=0$. It correspond to $k_x=\pm\pi,\,k_y=0$ in the BZ. Therefore the location of singular points depends on the choice of gauge.

The reason that any gauge has singular points is simply that the Chern number being nonzero implies no continuous gauge can cover the whole Brillouin zone.  It is easiest to understand the connection of these two by thinking of a spherical rather than toroidal Brillouin zone.  If a single gauge covered the whole sphere, then we could apply Stokes' theorem to relate the Chern number, which is the integral of the Berry curvature over the whole sphere, to the integral of the Berry connection around a tiny circle, which must be zero.  In the same way, a nonzero Chern number is an ``obstruction'' to having continuously defined wavefunctions over the whole Brillouin zone.

Each form of wave function is valid only on one patch of the torus and the two are connected by $U(1)$ gauge transformation on the boundaries. This is exactly the same as for the Wu-Yang monopole. $\psi_1^A$ and $\psi_1^B$ define a $U(1)$ bundle on the torus. Let $U^A$ be the open set which covers the torus without the point $k_x=k_y=0$.  Let $U^B$ be the open set which covers the torus without the point $k_x=\pm\pi,\,k_y=0$. Then $\psi_1^A$ is well defined on $U^A$ and $\psi_1^B$ is well defined on $U^B$. Both $\psi_1^A$ and $\psi_1^B$ are well defined on the overlap $U^A\bigcap U^B$ and they are related by $U(1)$ gauge transformation $\psi_1^A=e^{i\phi}\psi_1^B$. We can take a closed loop as the boundary of $U^A$ and $U^B$, then we find a mapping from this closed loop to $U(1)$. Then the Chern number is just the winding number of this mapping.

We can make this connection more explicit by taking a small closed loop as $c: \,k_x^2+k_y^2=\epsilon^2$ with small positive $\epsilon$ to enclose the singular point of $\psi_1^A$. Making a small $\vk$ expansion, we find $R_{1,2}\approx k_{x,y}/2$, $R_3\approx1$, $R\approx1$. Then we have
\be
\phi\approx-\arctan\frac{k_y}{k_x},\quad
\nabla_{\mu}\phi\approx\Big(\frac{k_y}{k_x^2+k_y^2},-\frac{k_x}{k_x^2+k_y^2}\Big)
\ee
Thus the Chern number can also be obtained as a line integral around this small loop
\be
c=\frac{1}{2\pi}\oint_c ({\bf A}^B-{\bf A}^A)\cdot d{\bf l}
=\frac{1}{2\pi}\oint_c \nabla_{\mu}\phi dl_{\mu}
=\frac{1}{2\pi}\oint_c\frac{k_ydk_x-k_xdk_y}{k_x^2+k_y^2}=-1
\ee
Even without detailed calculation, on can see that as the momentum goes around this loop, the phase difference $\phi$ of Eq (\ref{u1}) also goes around one circle. Thus the winding number is $-1$, which agrees with the previous calculation.

The Chern number can also be defined as a line integral $c=\frac{1}{2\pi}\oint_C A_{\mu}dl_{\mu}$, where integral path $C$ is the boundary of the BZ. Since the Berry vector potential depends on the gauge choice, to get a correct answer for the Chern number, one has to choose a gauge such that the wave function has no singular point inside the loop. In the Cu-O model, we should use $A^B_{\mu}$. It is easy to see that $A^B_x=0$ for $k_y=\pm\pi$ and $-\pi\le k_x\le \pi$ and $A^B_y=0$ for $k_x=\pm\pi$ and $-\pi\le k_y\le \pi$. Thus we find loop integral $\oint A_{\mu}dl_{\mu}=0$. But since there are two singular points $k_x=\pm\pi$, $k_y=0$ on the boundary, we should choose an integral path with a small semi circle to circumvent the two singular points. Expanding $A^B_{\mu}$ around $k_x=\pm\pi$, $k_y=0$, we find
\be
A^B_{\mu}(\pm\pi+k_x,k_y)=\frac{1}{k_x^2+k_y^2}\Big(k_x,-k_y\Big)
\ee
Let $\epsilon$ denote the small circle around the singular point, then
\be
c=\frac{1}{2\pi}\oint_C A_{\mu}dl_{\mu}=-\frac{1}{2\pi}\oint_{\epsilon}
\frac{k_xdk_y-k_ydk_x}{k_x^2+k_y^2}=-1
\ee
This result agrees with the Chern number calculated from the Berry curvature.

For the middle band $E=0$ and the wave function is $\ket{\psi_2}=\frac1R(R_3,R_2,R_1)$ which is always real in the BZ. Clearly this band is topologically trivial and Chern number $c=0$. The sum of the Chern numbers of all bands is always zero. Thus for the top band we have $c=1$.

Now we come back to the original Hamiltonian $H$. We show that if $t'$ is not too large, there is no band crossing so the result stays the same as $H'$. Let $R_4=ts_xs_y$, the eigenvalues are the roots of cubic equation
\be
E^3-(R_4^2+R^2)E+2R_4R_1R_2=0
\ee
The condition for degeneracy is $-[(R_4^2+R^2)/3]^3+(R_4R_1R_2)^2=0$, It can be rewritten as
$$
-[(\hat{R}_4^2+1)/3]^3+(\hat{R}_4\hat{R}_1\hat{R}_2)^2\le
-[(\hat{R}_4^2+1)/3]^3+\hat{R}_4^2/4=0
$$
with $\hat{R}_4=R_4/R$. The last equality of the above  is satisfied only for $\hat{R}_4^2=1/2$. In our model, for simplicity, we take $r=t$, then the maximum value of $\hat{R}_4$ is $\frac{t'}{2t}$ at $k_x=k_y=\pi$. Thus as long as $t<t'$, there is no band crossing and the Chern numbers stay the same as $H'$.

It follows from the discussion above that if the band is only partially filled, i.e. there is a Fermi surface, there can be no topologically protected currents. Nor can there be any singular properties of the Fermi liquid coming from the physics of the (non-quantized) anomalous Hall effect because one can always move the singularities of the wave-functions away from the Fermi surface by a suitable choice of gauge.   The intrinsic contribution to the anomalous Hall effect can be calculated by integrating the Berry curvature over the Fermi surface volume.  This procedure is best carried out numerically by writing the Chern number in terms of the projection operator~\cite{avronseilersimon} onto the occupied subspace, $P_k = |\psi_k \rangle \langle \psi_k|$, as this object is manifestly gauge-invariant.

\section{Conclusions and remarks on three-dimensional materials}

We have discussed the calculation of Berry phases and the contribution to the Hall effect in a three-band model motivated by the copper-oxygen planes of the high-temperature superconductors.  There are two topologically active bands (i.e., with nonzero Chern number $\pm 1$) separated by a middle band with Chern number zero.  We have neglected interaction effects, but there is considerable recent interest in the possibility of fractional quantum Hall phases when a band of nonzero Chern number is partly occupied.  Bands with Chern number $\pm 2$ or larger are particularly interesting as, while these are formally equivalent to bilayer quantum Hall systems, it has been argued that they are likely to support novel fractional quantum Hall states because the nature of interactions is modified~\cite{ranchern2}.  That work proposed creating the Chern number $\pm 2$ bands via an oxide heterostructure, and this technique might also enable the $PT$-breaking model described here to be realized at the interface between a cuprate and another material.

In closing, it may be useful to mention another topological property enabled when a model has three bands rather than two; the $CP$ violation enabled by the three-by-three Cabibbo-Kobayashi-Maskawa matrix in the Standard Model provides a well-known example of how three-by-three matrices allow additional subtleties.  The orbital contribution to the linear magnetoelectric effect, i.e., the polarization induced by a magnetic field $dP_i/dB_j$ or magnetization by an electric field $dM_i/dE_j$, has recently been a subject of active study~\cite{zhang,essinmoorevanderbilt,malashevich,essinturnermoorevanderbilt}.  The topological part of this effect is the scalar diagonal part (``axion electrodynamics''), computed by the Chern-Simons integral over the Berry connection~\cite{zhang}, which vanishes in a purely 2D model such as that considered here.

In a 3D model, either $P$ or $T$ symmetry quantizes the scalar diagonal part to only two possible values, corresponding to ordinary and topological insulators.  Without these symmetries, in a 3D model with only two bands, the Chern-Simons integral is quantized and computes the ``Hopf invariant'' of the band structure, viewed as a mapping to the sphere~\cite{hopfinsulator}.  Three bands are required in order to generate generic values of the scalar magnetoelectric coupling.  The computation for a single occupied band, in which case the Chern-Simons integral is Abelian, has been recently discussed~\cite{cohvanderbilt}.  We hope that the results of this paper lead to further study of the consequences of wavefunction topology for transport and magnetism of oxide materials.

\end{document}